\documentclass[aps,pre,twocolumn,floatfix,
nofootinbib,showpacs,longbibliography]{revtex4-1}

\usepackage{mathtools}
\usepackage{color}
\usepackage{color}
\usepackage{blkarray, bigstrut}
\usepackage{relsize}
\usepackage{amsmath}
\usepackage[british]{babel}  %Do hyphenation according to british english
\usepackage[scaled=1.03]{inconsolata} %Monospace font

\usepackage[colorlinks=true, citecolor=blue, urlcolor=blue]{hyperref}  %Hyperlinks (pink, green, blue)
\usepackage{graphicx} % Package to insert exteral figures
\usepackage[babel]{microtype}  %Improves text justification
\usepackage{amsmath,amssymb,amsthm,bm,amsfonts,mathrsfs,bbm} %Usefull math packages
\usepackage{xspace}  %Useful to add space in macros
\usepackage{pgfplots}
\usepackage{amsmath}
\usepackage{amssymb}

\newcommand{\ket}[1]{|#1\rangle}
\newcommand{\bra}[1]{\langle#1|}

\begin{document}

\title{Allowed and Forbidden Bipartite Correlations from Thermal States}

\author{Tamal Guha}
\email{g.tamal91@gmail.com}    
\affiliation{Physics and Applied Mathematics Unit, Indian Statistical Institute, 203 B.T. Road, Kolkata 700108, India.}

\author{Mir Alimuddin}
\email{aliphy80@gmail.com}    
\affiliation{Physics and Applied Mathematics Unit, Indian Statistical Institute, 203 B.T. Road, Kolkata 700108, India.}

\author{Preeti Parashar}
\email{parashar@isical.ac.in}    
\affiliation{Physics and Applied Mathematics Unit, 
	Indian Statistical Institute, 
	203 B.T. Road, Kolkata 700108, India.}

\begin{abstract}
	The strong connection between correlations and quantum thermodynamics raises a natural question about the preparation of correlated quantum states from two copies of a thermal qubit. In this work we study the specific forms of allowed and forbidden bipartite correlations. As a consequence, we extend the result to Separable (SEP) but not Absolutely Separable (AbSEP) class of product states. Preparation of a general form of entanglement from arbitrary thermal qubits is studied and as an application we propose a strategy to establish sustained entanglement between two distant parties. The threshold temperature to produce entanglement from two copies of a thermal qubit has also been discussed from the resource theoretic perspective, which ensures that the bound on the temperature can be superseded with the help of a resource state. A dimension dependent upper-bound on the temperature is derived, below which two copies of any $d-$dimensional thermal state can be entangled in $2\times d$ dimension.
\end{abstract}
\pacs{05.70.-a, 03.67.Bg, 03.67.-a}
%\keywords{}

% 03.65.Ta	Foundations of quantum mechanics;
% 03.67.Dd	Quantum cryptography and communication security
% 03.67.Hk	Quantum communication
% 03.65.Ud	Entanglement and quantum nonlocality

\maketitle
\section{Introduction}
Thermodynamics is one of the most well-understood and widely applicable traditional physics to analyze the macroscopic characteristics of several physical systems, viz., from the hydrostatic system to magnetizing material . Study of thermodynamics in the quantum regime has attracted attention to answer the validity of several thermodynamics laws and their features even in finite particle scenario. As a result, in the recent past, a lot of ideas regarding its basic foundation \cite{resource, secondlaw, apriori}, extraction of work \cite{popescuNAT, allahavardyan,alicki}, thermal machine \cite{franco_engine,refrigerator,chargingbattery,solarcellAlicki} etc. have enriched the subject in their own way.\par
Correlations among the constituents of a joint quantum system and their implication in information theory is deeply connected to the thermodynamics in quantum regime \cite{bera'NAT,bera'book}.
From the perspective of work extraction, the correlations present in quantum systems play an important role. One can significantly enhance the extraction of work by exploiting these correlations under the action of a global unitary. Such kind of advantage is primarily exhibited in \cite{horo'prl} for the case of degenerate Hamiltonian and subsequently studied in \cite{huberPRX, manikPRE, our}. Due to a well-established resource theory for quantum thermodynamics \cite{resource}, it is interesting to quantify the cost of creating such correlations, given some copies of free states. The amount of energy required to correlate two or more thermal states allowing fundamental limitations invoked by quantum theory have been studied in \cite{huber'njp}, along with the quantification of possible correlations given an amount of energy as resource.\par 
Although there are several fascinating examples where quantum correlations exhibit advantage over its classical counterpart viz., teleportation \cite{teleportation}, super-dense coding \cite{densecoding}, quantum key distribution \cite{ekart}, Bayesian games \cite{bayesian}, even over the PR correlation \cite{ALC}, all these tasks intrinsically depend on the particular form of correlation shared between the parties. So, it is of great interest to not only quantify the amount of correlations, but also specify the particular forms which can be created among the quantum systems.\par 
In this article we ask a similar question in the thermodynamic scenario, i.e., to specify the form of correlation which can be prepared from two thermal qubits kept at same or different temperatures. However for our analysis we shall ignore the cost of creating these allowed correlations. From practical point of view it is convenient to consider the marginals of the correlated state to be thermal in nature. This restriction forbids the preparation of a class of correlated states to be prepared from two copies of thermal qubits, even at different temperatures. It has already been studied in \cite{ghosh16,thermalbath} that quantum entanglement with thermal marginals can not be sustained for long if they are preserved in different temperature local baths. Motivated by this practical scenario we have designed a complete protocol to sustain a shared entanglement between two distant labs of different temperatures using a maximally entangled state as resource under LOCC. It is worth noticing that our result has a significant implication for product states of qubits. Although there is a class of separable states which can be entangled and conversely for every entangled state there exist several separable states from which it can be prepared, this both way connection is not true in general for product states. Here we have characterized several forms of entanglement which can not be prepared from two thermal or product qubits. However, for higher dimensional systems the inherent structure of quantum theory restricts us to draw a direct connection between the impossibility of producing entanglement from thermal and product states.\par 
Furthermore, Huber {\it et. al.} \cite{huber'njp} have shown that there is a temperature bound for qubit states, beyond which creation of entanglement from two copies of the same state, is impossible. In our work we go beyond this bound by raising the temperature of one qubit, while lowering the other. We have also extended this result to prepare a $2\times d$ dimensional entangled state, given two copies of a thermal qudit. We further cast and interpret our results for qubit state space in the framework of a resource theory. In general, a resource theory characterizes the potential to execute some specified task for any arbitrary state in comparison to those for which the resource amounts to zero. In recent past, several resource theories regarding entanglement \cite{res_ent, res_ent_1}, coherence \cite{res_coh}, thermodynamics \cite{resource}, purity \cite{res_pure}, quantum channels \cite{res_channel}, contextuality \cite{res_context} etc., have been developed. Here we define a set of free states such that none of their product state can be entangled under the action of global unitary. In that sense this is the resource theory for the constituents of separable (SEP) but not absolutely separable (AbSEP) product qubits.\par
The article is organized as follows. In Sec. II we discuss the necessary framework used in this work. In Sec III we study two different classes of forbidden correlations, namely classical correlations and entanglement. Sec IV contains the structure of the allowed entanglement and its application to prepare a sustained entanglement between two distant labs. In Sec V, we primarily go beyond the temperature limit for thermal qubits to make them entangled and then a general bound is derived for qudit case below which they can always be entangled and as a consequence a resource theory on the qubit state space is formulated depending upon their potential to produce entanglement. We finally conclude our results in Sec VI and the detailed calculations are given in the Appendix.
\section{Framework}
\subsection{Thermal State}
The state of a {\it d}-dimensional quantum system, governed by the Hamiltonian $H=\sum_{k=0}^{d-1}\epsilon_{k}|k\rangle \langle k|$ is said to be thermal {\it if and only if} it is the state of minimum energy with constant entropy or, maximum entropy with constant energy. For any state $\rho$, the internal energy and entropy of the system are defined as $E(\rho)=Tr(\rho H)$ and $S(\rho)=-Tr(\rho \log \rho)$ respectively.
The required thermal state takes the form \cite{lenard,skrzypzyckPRE} 
\begin{equation}
	\tau_{\beta}=\frac{e^{-\beta H}}{Tr(e^{-\beta H})}~~,
\end{equation}\label{thermal}
 where $\beta=\frac{1}{K_{B}T}$. Such a state at $\beta-$inverse temperature shall henceforth be called as $\beta-$thermal.
 \\%In our frame work for the possibilities of forming correlation we have chosen d=2.
 In this work, we will be mainly dealing with qubits. Our goal is to establish correlations between these thermal qubits by applying a global unitary on the joint thermal state. We consider the thermal states which are kept in the same as well as different temperature baths. The cost of forming these correlations is given by
 \begin{equation}\label{cost}
 W_{cost}=Tr(H_{total}\rho_{12})-Tr(H_{total}\tau_{\beta_{1}}\otimes\tau_{\beta_{2}})
 \end{equation}
 where $H_{total}=H_{1}\otimes I_{2}+I_{1}\otimes H_{2}$ is the total Hamiltonian and $\rho_{12}$ is the final correlated state.
\subsection{Classical Correlations}
We know that correlations in a bipartite quantum state can be characterized by Mutual Information (MI) between two parties defined as
\begin{equation}\label{mutual}
	I(A:B)=S(\rho_{A})+S(\rho_{B})-S(\rho_{AB}).
\end{equation} 
 Nonzero MI does not imply solely the presence of classical correlations. Although it quantifies the total correlations, it doesn't say anything about the nature or form of these correlations. As an example, the quantum states $\frac{1}{2}|00\rangle \langle 00|+\frac{1}{2}|11\rangle \langle 11|$ and $\frac{1}{2}|01\rangle \langle 01|+\frac{1}{2}|10\rangle \langle 10|$ possess completely different form of correlations, though they have the same amount of MI (=1).
\subsection{Entanglement}
To detect the entanglement for $2\times2$ and $2\times3$ dimensional systems, partial transpose is a necessary and sufficient criterion \cite{ppt'PLA}. A given state $\rho_{AB}$ will be entangled {\it if and only if} $(I_{A}\otimes\Lambda^{T_{B}})(\rho_{AB}) < 0$, where $\Lambda^{T_{B}}$ is the action of Transposition map on $B$. However, for higher dimensional systems this criterion is only sufficient. For pure bipartite quantum states, von-Neumann entropy of the reduced marginal is a good measure of entanglement \cite{ref}, whereas, for general mixed states concurrence is a widely used measure \cite{wooters'PRL}. However here we shall be focusing on the nature and not the amount of entanglement.
\subsection{Extension to SEP/AbSEP region}
A bipartite state $\rho_{AB}$ is said to be {\it separable} (SEP), {\it if and only if} it can be decomposed as $\sum_{i}p_{i}\rho_{A}^{(i)}\otimes\sigma_{B}^{(i)}$, where $\rho_{A}^{(i)}\in \mathcal{H}_{A}$ and $\sigma_{B}^{(i)}\in \mathcal{H}_{B}, \forall i$ and $p_{i}$ is the corresponding probability.
\\ A refinement on the set of separable states was introduced in \cite{marek,knill} depending upon their spectrum. A bipartite separable state $\rho_{AB}$ is said to be absolutely separable (AbSEP) {\it if and only if} the state remains separable even after application of all possible global unitary $\mathcal{U}\in \mathcal{L}(\mathcal{H}_{A}\otimes\mathcal{H}_{B})$. A significant amount of study has been done to characterize the set AbSEP for $2\times d$ systems \cite{Johnston'13}.
\\Besides the characterization of these AbSEP states, another interesting question which can be asked is whether there are some entangled states which can't be prepared from the separable but not absolutely separable states? Although the answer is in general negative, one can modify the search by considering only product states from this region. In the present work, we have investigated and characterized several classes of forbidden and allowed entangled states from two-qubit product states belonging to the SEP but not AbSEP region, i.e., SEP/AbSEP region. 
\subsection{Limits on thermal states for creating correlation}
It is interesting to ask whether any two thermal qubits can be correlated under the action of global unitary. It is trivial that the product of two pure states or two maximally mixed states are the only forbidden elements to produce classical correlations. However, the scenario is not so simple for the question of entanglement production. Concurrence \cite{wooters'PRL} is a good measure to detect the entanglement present in mixed bipartite quantum states, which is a spectrum-dependent criterion. Now, as the action of global unitary preserves the spectrum of joint state, the same expression for concurrence can restrict the spectrum of initial qubit to be entangled. Incorporating this fact in \cite{huber'njp} a {\it threshold temperature} has been derived, i.e., $\frac{k_BT}{E} < 1.19$, beyond which two copies of the same thermal qubit can not be entangled applying global unitary.

\section{Forbidden Correlations}
Before discussing the forbidden correlations, we will try to formally classify all kinds of non-trivial \cite{comment} correlations which can be prepared from two thermal states of same temperature, taking them as free resources.\\
\\
{\bf Theorem 1 :~~}{\it If a two-qubit correlated state $\rho$ can be prepared from two same temperature thermal states under the action of global unitary, then $\rho$ must has a degenrate eigenvalue with degeneracy 2, and the eigenvalues are in GP.}\\
\\{\bf Proof:~~} Suppose the bipartite correlated state $\rho$ can be prepared from two copies of a single qubit state $\sigma$. Let the eigenvalues of $\rho$ be $\lambda_1\ge\lambda_2\ge\lambda_3\ge\lambda_4$ and those of $\sigma$ be $\delta\ge(1-\delta)$. Then the eigenvalues for $\sigma^{\otimes2}$ would be $\delta^2\ge\delta(1-\delta)\ge(1-\delta)^2$, where $\delta(1-\delta)$ is a degenerate eigenvalue with degeneracy 2.
The unitary action preserves the spectrum of initial state, and therefore to obtain $\rho$ from $\sigma^{\otimes2}$, two intermediate eigenvalues of $\rho$ should be same. Hence, $\lambda_2=\lambda_3$.
\\Comparing the eigenvalues of these states, gives $\lambda_1=\delta^2$, $\lambda_2=\delta(1-\delta)$ and $\lambda_4=(1-\delta)^2$. Therefore, $\lambda_1 \lambda_4= \lambda_2^2$, i.e., $\lambda_1,\lambda_2$ and $\lambda_4$ are in Geometric Progression (GP).$~~~~\blacksquare$

\subsection{Impossibility in classical correlation}
We now consider some bipartite quantum states with local marginals as $\beta-$thermal. The preservation of the spectrum under the action of global unitary forbids a class of classically correlated states to be prepared from the two same (or, different) temperature thermal qubits. Spectrum preservation sufficiently implies that entropy of the final joint state $\rho_{AB}^{f}$ would be equal to that of the  initial product state $\sigma_{A}^{i}\otimes\sigma_{B}^{i}$, i.e.,
\begin{equation}\label{entropy}
\begin{aligned}
 S(\sigma_{A}^{i})+S(\sigma_{B}^{i})&=S(\rho_{AB}^{f}) \\
 S(\sigma_{A}^{i})+S(\sigma_{B}^{i})&=S(\rho_{A}^{f})+S(\rho_{B}^{f})-I_{\rho}(A:B) 
 \end{aligned}
\end{equation}
\\In order to obtain a classically correlated final states the mutual information $I_{\rho}(A:B)$ should be grater than zero, which implies that $S(\rho_{A}^{f})+S(\rho_{B}^{f})>S(\sigma_{A}^{i})+S(\sigma_{B}^{i})$. We know that entropy is a monotonically increasing function of temperature for a qubit \cite{bera'book}. Since the total marginal entropy should increase, it is in no way possible to reduce the temperature of both the initial qubits simultaneously. This means that we can only prepare those states for which the sum of the final local temperatures is higher than the sum of the initial local temperatures. Thus some of the classical correlations are intrinsically forbidden. In the following we shall consider various classes of classically correlated states and show their impossibility from same or different temperature thermal states. We assume that the final marginals are at the same temperature, i.e., $\rho_{A}^{f}=\rho_{B}^{f}$.
\\
\\{\bf Case-1:}
\\Let's consider the state
\begin{equation}\label{cct}
\rho_{AB}= p \frac{|00\rangle\langle00|+e^{-\beta E}|11\rangle\langle11|}{Z} + (1-p) \tau_{\beta}\otimes\tau_{\beta}
\end{equation}
where $\tau_{\beta}=\frac{|0\rangle\langle0|+e^{-\beta E}|1\rangle\langle1|}{Z}$ is a $\beta-$thermal state with the partition function $Z=1+e^{-\beta E}$, and  governed by the local Hamiltonian $H_{A}=H_{B}=E|1\rangle\langle1|$.
 \\$\rho_{AB}$ possesses classical correlations (non-zero mutual information) for $p\in (0,1]$ and it's eigenvalues are $\lambda_1=\frac{(p+e^{\beta E})e^{\beta E}}{(1+e^{\beta E})^2},~~~  \lambda_2=\lambda_3=\frac{(1-p)e^{\beta E}}{(1+e^{\beta E})^2}$ and $\lambda_4=\frac{(1+pe^{\beta E})}{(1+e^{\beta E})^2}$. For the region $p\le\frac{1-e^{-\beta E}}{2}$, they follow the non-increasing order $\lambda_1\ge\lambda_2=\lambda_3\ge\lambda_4$ %(see fig.1, with $\beta E=q$). 
 %\begin{figure}[htb!]
 %	\includegraphics[scale=0.5]{Case_1.png}
 %	\caption{ The \textcolor{teal}{thick meshed} plane represents $\lambda_1$. Similarly, $\lambda_{2}=\lambda_{3}$ is denoted by \textcolor{blue}{mesh-free} plane and $\lambda_{4}$ by \textcolor{red}{dashed} plane.}
 %\end{figure}
 
 \iffalse
 Now, if we want to prepare this correlation from two different thermal states $\tau_{\beta_{1}}$ and $\tau_{\beta_{2}}$ with spectrum {s,1-s} and {t,1-t} respectively with s,t$\ge\frac{1}{2}$, then the equality of $\lambda_2$ and $\lambda_3$ demands s=t, for some nontrivial \cite{comment} solution. This yields the condition $e^{\beta E}=-1$, which is impossible. So, it is impossible to prepare the correlated state of the form (4) from two product thermal states.
 \fi
Now, Theorem 1 suggests that the state $\rho_{AB}$ can be prepared from two same temperature thermal states if the eigenvalues are in G.P. i.e., $\lambda_{1}\lambda_{4}=\lambda_{2}^{2}$. This yields that $\cosh(\beta E)=-1$, which is impossible.\par
For the complimentary region of $p$ and $\beta E$, where $\lambda_{1}\ge\lambda_{4}\ge\lambda_{2}=\lambda_{3}$, {\it either} the state (\ref{cct}) with $p=1$ can be prepared from two different thermal states of temperatures $\frac{1}{\beta E}$ and zero, {\it or} from $\frac{\mathbb{I}}{2}$ and a $\ln(\frac{1+p}{1-p})-$thermal state only for $\beta\to 0$. The detailed analysis can be found in the {\it Appendix}.\\
 \\{\bf Case-2:}
 \\Here we will consider the state
 \begin{equation}\label{entt}
 \rho_{AB}=p\tilde{\Phi}+(1-p)\tilde{\Psi}
 \end{equation}
 where, $\tilde{\Phi} = \frac{|00\rangle\langle00|+e^{-\beta E}|11\rangle\langle11|}{Z}$ \\and   $~~~~\tilde{\Psi} =\frac{|01\rangle\langle 01|+e^{-\beta E}|10\rangle\langle 10|}{Z}$.
 \\The temperature of local marginals for the state (\ref{entt}) will be same only for p=1. Otherwise both the marginals can be thermal but with different temperatures for $p\ge\frac{1}{2}$. In this region, eigenvalues corresponding to the joint state will be  $\lambda_1=\frac{p e^{\beta E}}{1+e^{\beta E}}, \lambda_2=\frac{(1-p)e^{\beta E}}{1+e^{\beta E}}, \lambda_3=\frac{p}{1+e^{\beta E}}$ and $\lambda_4=\frac{(1-p)}{1+e^{\beta E}}$, where $\lambda_1$ and $\lambda_4$ are maximum and minimum respectively. But the order between $\lambda_2$ and $\lambda_3$ can alternate depending upon p and $\beta E$. For $p\le \frac{1}{1+e^{-\beta E}}$ the ordering is 
 $\lambda_1\ge\lambda_2\ge\lambda_3\ge\lambda_4$ while for $p\ge \frac{1}{1+e^{-\beta E}}$ it is $\lambda_1\ge\lambda_3\ge\lambda_2\ge\lambda_4$.
  %\begin{figure}[htb!]
  %	\begin{center}
  %	\includegraphics[scale=0.5]{Case_2.png}
  %	\caption{The \textcolor{brown}{thick meshed} plane represents $\lambda_1$. Similarly, $\lambda_{2},\lambda_{3}$ and $\lambda_{4}$ is denoted by \textcolor{red}{dashed}, \textcolor{blue}{mesh-free} and \textcolor{teal}{striped} planes respectively.}
  %\end{center}
  %\end{figure}
  For both regions one can have two different initial thermal states of temperatures $\beta$ and $\ln(\frac{p}{1-p})$ to achieve this correlation. It is worth noticing that to create the correlation (\ref{entt}) in general one should choose two different initial thermal states. Although the state has classical correlation for $p\in[0,1]$, it is possible to prepare the state from two same initial thermal states for $p=\frac{1}{1+e^{-\beta E}}$ only.
 \\
 \\
 \\{\bf CASE-3:}
 \\Here, we consider another forbidden classically correlated state of the form
  \begin{equation}\label{forbidden}
 \rho_{AB}=p\tilde{\Phi}+q\tau_{\beta}\otimes\tau_{\beta}+(1-p-q)|\tilde{\phi}^+\rangle\langle\tilde{\phi}^+|
  \end{equation}
  where $|\tilde{\phi}^+\rangle=\frac{|00\rangle+e^{\frac{-\beta E}{2}}|11\rangle}{\sqrt{Z}}$, with other terms as defined earlier. Due to the presence of a non-maximally pure entangled component, the NPT criterion \cite{ppt'PLA} reveals that the state will be separable with non-zero MI for $\cosh(\frac{\beta E}{2})\le \frac{q}{2(1-p-q)}$.
  %\begin{figure}[htb!]
  %	\begin{center}
  %		\includegraphics[scale=0.5]{Case_3.png}
  %		\caption{In the \textcolor{brown}{mesh-free} region $\lambda_{-}\ge \lambda_{2}=\lambda_{3}$. Similarly, \textcolor{teal}{meshed} part depicts $\lambda_{-}\le \lambda_{2}=\lambda_{3}$. Both the region is plotted under the condition $\cosh(\frac{\beta E}{2})\le \frac{q}{2(1-p-q)}$ }
  %	\end{center}
  %\end{figure}
\\
  In this region the eigenvalues corresponding to the state (\ref{forbidden}) can be ordered as $\lambda_1=\lambda_+\ge\lambda_2=\lambda_3\ge\lambda_4=\lambda_-$ for one part and $\lambda_1=\lambda_+\ge\lambda_4=\lambda_-\ge\lambda_2=\lambda_3$ for the other part where, $\lambda_2=\lambda_3=\frac{qe^{-\beta E}}{(1+e^{-\beta E})^2}$ and
  $\lambda_{\pm}=\frac{1+2e^{-\beta E}(1-q)+e^{-2\beta E}\pm(1+e^{-\beta E})\sqrt{1+(4(1-p-q)^2-2)e^{-\beta E}+e^{-2\beta E}}}{2(1+e^{-\beta E})^2}$.
  For the first region the equality of $\lambda_2$ and $\lambda_3$ depicts that the initial product states should be of same temperature to produce this correlation. As a result, $\lambda_2$ should be the geometric mean of $\lambda_1$ and $\lambda_4$, which is only possible for $\beta\to\infty$, which is the trivial solution, with no correlation in the final state. For the second region where the lowest two eigenvalues are degenerate, we have to choose product of two different temperature thermal states, to make them correlated. In this case one of the two initial product states will be trivial.
  
  \subsection{Impossibility in Entanglement}
  In this section we will focus on the forbidden as well as allowed entangled states prepared from two product thermal states. Before describing the several classes of entangled states which are forbidden from two identical (or different) temperature thermal qubits, we shall introduce a mathematical result relating this impossibility to the class of SEP but not AbSEP product states.\\
\\
{\bf Theorem 2 :~~}{\it Any two-qubit $2\times 2$ entangled state $\rho$ can be prepared from a} SEP but not AbSEP {\it product state, under the action of global unitary, iff $\rho$ can be prepared from the product of two thermal qubits.}\\
{\bf Proof:} The {\it if} part is trivial. We will consider the {\it only if} part here. Suppose there is a bipartite state $\rho$ which can't be prepared from the product of any two thermal states, but can be prepared from the product state $\sigma\otimes\xi$ under the action of global unitary $\mathcal{V}$. However in the case of qubits trivial constraints on probability demands that it is always possible to find two thermal states $\tau_{\beta_1}$ and $\tau_{\beta_2}$ having same spectrum as $\sigma$ and $\xi$ respectively. Therefore there exists some local unitary
 $\mathcal{U}_1 \otimes \mathcal{U}_2$, such that $\mathcal{U}_1(\tau_{\beta_1})\to\sigma$ and $\mathcal{U}_2(\tau_{\beta_2})\to\xi$.
\\ Thus the global action of $\mathcal{V}\circ(\mathcal{U}_1\otimes\mathcal{U}_2)$ on $\tau_{\beta_1}\otimes\tau_{\beta_2}$ will produce the desired state $\rho$, which is a contradiction to the above assumption.
$~~~~\blacksquare$\\
\\
Below we will characterize several classes of entangled states which are forbidden from two same (or different) temperature thermal qubits and try to construct a general kind of allowed entangled state.
\\
\\{\bf CASE-4:}
First we consider a general kind of Werner state with $\beta-$thermal marginal, such that for $\beta\to0$ the state becomes a variant of the well known Werner state given by
  \begin{equation}\label{werner}
  \rho_{AB}=p|\tilde{\phi}^+\rangle\langle\tilde{\phi}^+|+(1-p)\tau_{\beta}\otimes\tau_{\beta}
  \end{equation}
  where $|\tilde{\phi}^+\rangle$ is defined earlier. Taking the partial transpose with respect to B for (\ref{werner}) it can be shown that the state will be entangled for $p\ge\frac{1}{1+2\cosh(\frac{\beta E}{2})}$.
  \\Eigenvalues corresponding to the state (\ref{werner}) are given by, $\lambda_+\ge\lambda_2=\lambda_3\ge\lambda_-$, where $\lambda_{\pm}=\frac{1+2pe^{\beta E}+e^{2\beta E}\pm(1+e^{\beta E})\sqrt{(1-e^{\beta E})^{2}+4p^2e^{\beta E}}}{2(1+e^{\beta E})^2}$ and $\lambda_2=\lambda_3= \frac{e^{\beta E}(1-p)}{(1+e^{\beta E})^2}$.
   %\begin{figure}[htb!]
  	%\begin{center}
  	%	\includegraphics[scale=0.3]{Case_4.png}
  	%	\caption{\textcolor{blue}{mesh-free} plane represents $\lambda_{+}$. Similarly, \textcolor{teal}{striped} and \textcolor{red}{dashed} planes depict $\lambda_{2}=\lambda_{3}$ and $\lambda_{-}$ respectively.}
  	%\end{center}
  %\end{figure}
  \\The degeneracy in eigenvalues demands that this correlation can only be prepared (if possible) from two copies of same thermal states as a consequence of {\it Theorem 1}. The GP condition on the eigenvalues yields 
   \begin{eqnarray}\label{gp}
   \frac{1}{16}(1-p)(1+p+2p C(\beta E)) S^4(\frac{\beta E}{2})\nonumber\\
   =\frac{1}{16}(1-p)^2 S^4(\frac{\beta E}{2})
   \end{eqnarray}
   where, $S\equiv~sech$, and $C\equiv~cosh$.
   \\Eq.(\ref{gp}) demands either p=1, which is a trivial condition, or $sech^4(\frac{\beta E}{4})=0 \implies \beta E \to \infty$ which is again a trivial solution, or $cosh(\beta E)=-1$, which is impossible.
   \\
   \\{\bf CASE-5:}
   Here we consider the state,
     \begin{equation}\label{pmt}
     \rho_{AB}=p|\tilde{\phi}^+\rangle\langle\tilde{\phi}^+|+q|\tilde{\phi}^-\rangle\langle\tilde{\phi}^-|+(1-p-q)\tau_{\beta}\otimes\tau_{\beta}
     \end{equation} 
     Taking the partial transpose of (\ref{pmt}) with respect to B, it can be shown that the state will be entangled for $(1+p+q)e^{\frac{\beta E}{2}}\mp p(1\pm e^{\frac{\beta E}{2}})^2 \pm q(1\mp e^{\frac{\beta E}{2}})^2\le0$. 
     \\The eigenvalues of state (\ref{pmt}) are\\
      $\lambda_{\pm}=\frac{1}{2}-\frac{2(1-p-q)e^{\beta E}\pm(1+e^{\beta E})\sqrt{(e^{\beta E}-1)^2+4e^{\beta E}(p-q)^2}}{2(1+e^{\beta E})^2}$, and\\
      $\lambda2=\lambda3=\frac{(1-p-r)e^{\beta E}}{(1+e^{\beta E})^2}$, with $\lambda_+\ge \lambda_2=\lambda_3\ge\lambda_-$. {\it Theorem 1} demands that these states can only be prepared from two copies of same thermal state, if possible. For the existence of such a state, $\lambda_2$ should be the geometric mean of $\lambda_\pm$, which yields the condition\\
      $\frac{1}{4}[(p+q)-(p-q)^2] sech^2(\frac{\beta E}{2})=0\implies$ either, $\beta E\to\pm\infty$ (trivial solution), or, $(p+q)=(p-q)^2$, which is impossible. 
      
       \section{Allowed Entanglement}
       \subsection{Entangled State with Thermal Marginal}
     In the above, we have discussed several forms of forbidden correlations from two thermal states. Now we will focus on the general form of the allowed entanglement which can be created from two copies of a $\beta-$thermal state. The final marginal thermal states are at a higher temperature than the initial ones, which follows from the second law of thermodynamics \cite{secondlaw,bera'NAT}.\par
      Let us consider the $\beta-$thermal state $\tau_{\beta}=p|0\rangle\langle0|+(1-p)|1\rangle\langle1|$, where $p=\frac{1}{1+e^{-\beta E}}\ge\frac{1}{2}$. The spectrum of $\tau_{\beta}^{\otimes2}$ is $\{p^2, p(1-p),p(1-p),(1-p)^2\}$. An appropriate unitary can be chosen depending upon the single parameter $\theta$, such that $\tau_{\beta}^{\otimes2}\xrightarrow{\mathcal{U}_{\theta}}\rho_{AB}$. The explicit action of $\mathcal{U}$ on the energy eigen-basis of $\tau_{\beta}^{\otimes 2}$ is 
      $\mathcal{U}_{\theta}|k\rangle\to|\phi_{k}\rangle,~~k\in\{0,1\}^2$, where $|\phi_{00}\rangle=a|00\rangle+b|11\rangle,~~ |\phi_{11}\rangle=b|00\rangle-a|11\rangle,~~ |\phi_{01}\rangle=a|01\rangle+b|10\rangle$ and $|\phi_{10}\rangle=b|01\rangle-a|10\rangle$, with $a=\frac{1}{\sqrt{1+e^{-\theta}}}$ and $b=\frac{e^{\frac{-\theta}{2}}}{\sqrt{1+e^{-\theta}}}$.\\
      A simple calculation shows that the marginal temperature corresponding to the final entangled state can be characterized as $\beta^{'}E=\ln[\frac{(1-p)^{2} e^{-\theta}+p}{p^{2}e^{-\theta}+(1-p)}]$.\par
      
       \begin{figure}[!htb]
      	\begin{center}
      		\includegraphics[scale=0.5]{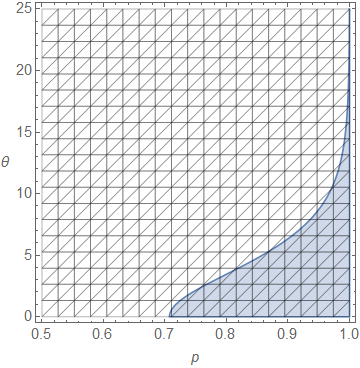}
      		\caption{Meshed region indicates that $\forall p \in [\frac{1}{2},1]$ and $\forall \theta \ge 0$ we have $ \beta^{'}E\ge \beta^{''}E$. However the final correlated state will only be entangled for the \textcolor{blue}{solid filled} region.}
      		\label{fig1}
      	\end{center}
      \end{figure}
     
      Although, there are $24$ possible permutations of the unitary, only 6 of them, which take $|00\rangle \to|\phi_{00}\rangle$, will create correlations such that the marginals are thermal. Among these $6$, only $2$ permutations with the action $|11\rangle \to|\phi_{11}\rangle$, will produce entangled states with same local temperature (quantified by $\beta^{'}E$). For the other $4$ permutations, the marginals would be characterized by different temperatures $\beta^{'}E$ and $\beta^{''}E=\ln[\frac{(1-p)^{2}+p^{2}+ 2p(1-p)e^{-\theta}}{[p^{2}+(1-p)^{2}]e^{-\theta}+2p(1-p)}].$

      \subsection{Sustained Entanglement}
    In this section we will focus on the practical scenario for quantum communication schemes. Quantum Teleportation \cite{teleportation} is one of the most fundamental information theoretic tasks between two distant parties which is permitted by the statespace structure of quantum mechanics itself. Let us consider that two distant parties, Alice and Bob, share a maximally entangled state $|\psi^{-}\rangle$. They can teleport a quantum state using this correlation locally along with 2-cbits. But to preserve the correlations in $|\psi^{-}\rangle$, the marginals are required to be kept in contact with infinite temperature thermal baths. However, from practical point of view, it is impossible to achieve this temperature. So instead, let the local baths be kept at inverse temperatures $\beta_{1}$ and $\beta_{2}$. Now by the following protocol, Alice and Bob can prepare a non-maximally entangled state and use it in future communication tasks, whose local marginals are saturated with their corresponding local bath, on the expense of the shared maximally entangled state.\\
    {\it Step 1:} Alice takes two $\beta_{1}$-thermal qubit from her bath and cools them down to the states of inverse temperature $\beta^{'}$. 
    \\Thermodynamic cost corresponding to this stimulated cooling will be same as the change in free energy  $F[\tau_{\beta^{'}}^{\otimes2}]-F[\tau_{\beta_{1}}^{\otimes2}]$ \cite{HuberPRE'15}. 
    \\{\it Step 2:} Alice will apply a proper unitary $\mathcal{U}(\theta)$ as described in the last section, to obtain an entangled state with $\beta_{1}-$thermal and $\beta_{2}-$thermal local marginals. The thermodynamic cost in this case can similarly be characterized as, $E[\tau_{\beta_{1}}\otimes\tau_{\beta_{2}}]-E[\tau_{\beta^{'}}^{\otimes2}]$. In this stage, the operation is executed under unitary evolution on the complete system, which guarantees that the change in free energy is equivalent to the change in total energy only.
    \\{\it Step 3:} Alice will utilize the pre-shared maximally entangled state $|\psi^{-}\rangle$ to teleport the $\beta_{2}-$thermal particle to Bob. As a consequence they can finally share a sustainable mixed entangled state between them for future communication. \\
     \begin{figure}[!htb]
    	\begin{center}
    		\includegraphics[scale=0.5]{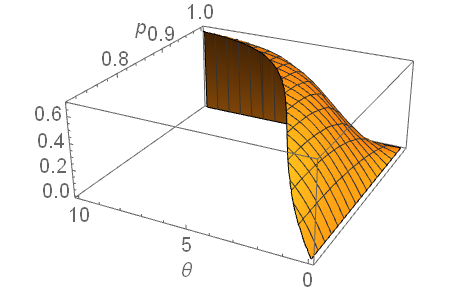}
    		\caption{The difference between $\beta_{1}E$ and $\beta_{2}E$ is plotted against $p=\frac{1}{1+e^{-\beta^{'}E}}$ and $\theta$. The surface depicts that there is an asymptotic upper bound ($\sim 0.693$) on the difference.}
    		\label{fig 2}
    	\end{center}
    \end{figure}
    Although, the full range of $\beta^{'}$ and $\theta$ are allowed to make the transformation, difference between the local temperatures is upper bounded by a finite number, explicitly, $(\beta_{1}-\beta_{2})E\le0.693$ ({\it see Fig. \ref{fig 2}})
       
      \section{Temperature Limit of Creating Entanglement}
      \subsection{The temperature bound}
        A separable state in $\mathbb{C}^{2}\otimes \mathbb{C}^{d}$ with eigen spectrum $\{\lambda_{1}\ge\lambda_{2}\ge...\ge \lambda_{2d}\}$ will remain separable under the action of all possible global unitaries, {\it if and only if} \cite{Johnston'13}
        \begin{equation}\label{abs}
        \lambda_{1}\le \lambda_{2d-1}+2\sqrt{\lambda_{2d-2}\lambda_{2d}}.
        \end{equation}
        
\begin{widetext}   	      	
       	
       	\begin{figure}[!b]
        \includegraphics[scale=0.5]{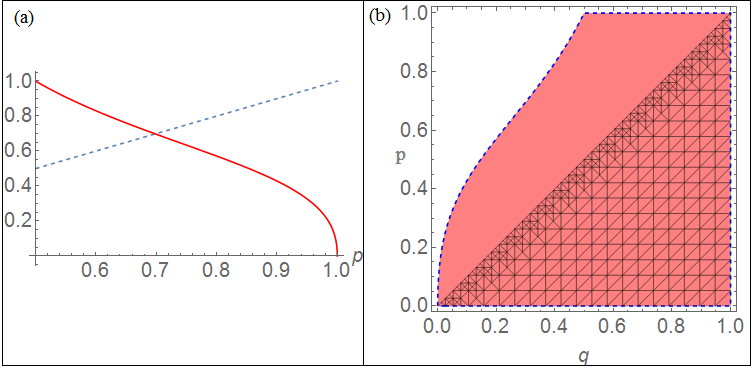}
        \caption{\textbf{(a)} Any value of $q$ lying above the \textcolor{red}{curve} can be taken along with $p$ to make their product entangled. Similarly, the \textcolor{blue}{dashed straight line} denotes $q=p$. So, there is no $q\le p$ for $p\le0.698$ to make their product entangled.~~\textbf{(b)} The meshed region stands for $p\le q$, whereas \textcolor{purple}{filled} area denotes allowed $p$ values for which their product can be entangled.}
        \label{fig3}
      \end{figure}  
    
  	    \end{widetext}

        Now, let's consider two copies of a given thermal state $\tau_{\beta} = \frac{1}{1+e^{-\beta E}}|0\rangle\langle0|+\frac{e^{-\beta E}}{1+e^{-\beta E}}|1\rangle\langle1|.$ In order to create entanglement, the eigenvalues of $\tau_{\beta}^{\otimes 2}$ will give a bound on $\beta E$, so that Eq.(\ref{abs}) can be violated. In \cite{huber'njp} the authors have shown that, for $\beta E\le0.84$, i.e., $\frac{K_{B}T_{max}}{E}\ge1.19$, no entanglement can be created from $\tau_{\beta}^{\otimes 2}$.
        \\ Here, we will show that it is possible to go beyond the temperature bound $T > T_{max}$ by taking another state with $T^{'} < T_{max}$. We consider two different thermal states $\tau_{\beta}=p|0\rangle\langle0|+(1-p)|1\rangle\langle1|$ and $\tau_{\beta^{'}}=q|0\rangle\langle0|+(1-q)|1\rangle\langle1|$. \\
        If $p\ge q$, then the eigenvalues of $\tau_{\beta}\otimes\tau_{\beta^{'}}$ can be written as $pq\ge p(1-q)\ge q(1-p)\ge (1-p)(1-q)$, which violates Eq.(\ref{abs}) for $q\ge \frac{2\sqrt{p(1-p)}}{2p-1+2\sqrt{p(1-p)}}$. This bound on $q$ is satisfied for only $p\ge0.698$ (since $p\ge q$ in this case) which exactly matches with the one obtained in \cite{huber'njp} (see {\it Fig. \ref{fig3}}).\\
       
        On the other hand, if $p\le q$, the eigenvalues can be arranged as,   $pq\ge q(1-p)\ge p(1-q)\ge (1-p)(1-q)$. In order to violate Eq.(\ref{abs}),\\
       $$pq>p(1-q)+2(1-p)\sqrt{q(1-q)}$$\\
        \begin{equation}\label{resource}
        	\implies \frac{(1-2q)}{\sqrt{q(1-q)}}\le\frac{2(1-p)}{p}.
        \end{equation}
        This means that for every $p\in[\frac{1}{2},1]$, there exists a $q$, where $1\ge q\ge p$, such that $\tau_{\beta}(p)\otimes\tau_{\beta^{'}}(q)$ can always be entangled under the action of global unitary.\\
    
      Another interesting question is to find the optimal temperature for higher dimensional quantum states, beyond which it is not possible to create entanglement. Although it is hard to find such a temperature, we give a dimension-dependent bound, below which any state would necessarily be entangled.\par
        Let us consider a {\it d} dimensional thermal state $\tau_{\beta}=\frac{e^{-\beta H}}{Tr(e^{-\beta H})}$, with $H=\sum_{k=0}^{d-1}\epsilon_{k}|k\rangle \langle k|$, where $\epsilon_{k}=kE$. One part of the initial state $\tau_{\beta}^{\otimes 2}\in \mathbb{C}^{d}\otimes\mathbb{C}^{d}$ is projected onto the subspace spanned by $\{|0\rangle, |d-1\rangle\}$. Then, $\tau_{\beta}^{\otimes 2}\to \tau_{\beta}\otimes\tau_{\beta}^{'}$, where,\\ $\tau_{\beta}^{'}=\frac{1}{1+e^{-(d-1)\beta E}}|0\rangle\langle0|+\frac{e^{-(d-1)\beta E}}{1+e^{-(d-1)\beta E}}|d-1\rangle\langle d-1|$.\\
        For $d\ge3$ the eigenspectrum corresponding to $\tau_{\beta}\otimes\tau_{\beta}^{'}$ will be,
         $\lambda_{1}=\frac{1}{ZZ^{'}}\ge ... \ge\lambda_{2d-2}=\frac{e^{-2(d-2)\beta E}}{ZZ^{'}}\ge\lambda_{2d-1}=\frac{e^{-(2d-3)\beta E}}{ZZ^{'}}\ge\lambda_{2d}=\frac{e^{-2(d-1)\beta E}}{ZZ^{'}}$
          where $Z=Tr(e^{-\beta H})$ and $Z^{'}=1+e^{-\beta E}$.\\
        \begin{figure}[!htb]
        	\includegraphics[scale=0.5]{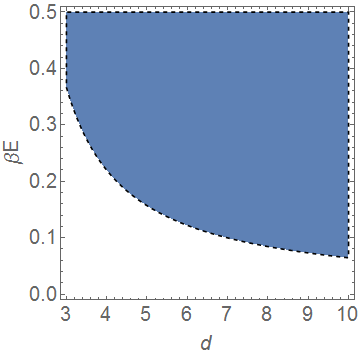}
        	\caption{The \textcolor{blue}{filled} region depicts the allowed values for $\beta E$ to create entanglement depending upon the dimension of the thermal state. It is possible to make the forbidden region for $\beta E$ arbitrarily small for asymptotically large {\it d}.}
        	\label{fig4}
        \end{figure}
        Now, according to Eq.(\ref{abs}), $\tau_{\beta}\otimes\tau_{\beta^{'}}$ can be entangled {\it if and only if}
        \begin{equation}
        \begin{aligned}
        1&\ge e^{-(2d-3)\beta E}+ 2\sqrt{e^{-2(d-2)\beta E}\times e^{-2(d-1)\beta E}} \nonumber \\
        \implies 1& \ge 3e^{-(2d-3)\beta E} \nonumber \\
        \implies\beta E & \ge \frac{\ln 3}{(2d-3)}\\
        \implies T & \le \frac{(2d-3) E}{K_{B} \ln 3}
         \end{aligned}
        \end{equation}
         The above criterion depicts that asymptotically for very large values of {\it d}, $\tau_{\beta}^{\otimes 2}$ can be entangled in $\mathbb{C}^{d}\otimes\mathbb{C}^{2}$ for arbitrarily small values of $\beta E$ ({\it see Fig. \ref{fig4}}). However, the optimality of this bound can not be guaranteed. \\
         \subsection{Resource theoretic framework}
         The above discussions open up the possibilities to form a resource theory on the temperature of thermal qubits. The bound introduced in \cite{huber'njp} demands that the states with temperature higher than the critical one can not be entangled using two copies, under the action of global unitary. So for the task of creating entanglement from two copies of thermal states under the action of global unitary, these states are free to access. More precisely, the set of free states is $\mathcal{F}:=\{\tau_{\beta}\in \mathbb{C}^{2}|\tau_{\beta}^{\otimes 2}\in AbSEP \}$. As a consequence of {\it Theorem 2}, the set of free states can be extended up to the product states connected by the unitary on these $\tau_{\beta}\in \mathcal{F}$. Under all possible global unitary, $\mathcal{U}_{g}(\rho\otimes\sigma)\to AbSEP, \forall\rho, \sigma \in \mathcal{F}$. We define $\mathcal{\bar{F}}$ as the set of all resource states having potential to create entanglement from two copies under unitary operations. The set $\mathcal{\bar{F}}$ can also be extended to the set of qubits for which product of any two lies in  the SEP/AbSEP region.\par
                    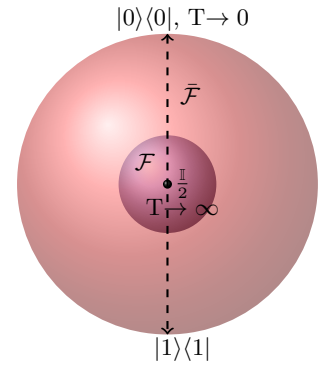
\begin{figure}[htb]
         	\scalebox{1.}{
         		\begin{tikzpicture}
         		\shade[ball color = blue!30, opacity = 1.0] (0,0) circle (0.65cm);
         		\shade[ball color = red!80, opacity = 0.5] (0,0) circle (2cm);
         		\node at (-0.3,0.3) {$\mathcal{F}$};
         		\node at (0.3,1.2) {$\mathcal{\bar{F}}$};
         		\shade[ball color=black] (0,0) circle (.06cm);
         		\draw[<->,dashed,thick](0,2)--(0,0)--(0,-2);
         		\node at (0.2,0.01) {$\frac{\mathbb{I}}{2}$};
         		\node at (0.2,-.3) {T$\to\infty$};
         		\node at (0.2,2.2) {$\ket{0}\bra{0}$, 	T$\to 0$};
         		\node at (0.2,-2.2) {$\ket{1}\bra{1}$};
         		\end{tikzpicture}}
         	\caption{(Color on-line) The external sphere denotes the qubit Bloch sphere and the inscribed blue ball represents the set $\mathcal{F}$. The annular region captures all the resource states $\mathcal{\bar{F}}$.}
         	\label{fig5}
         \end{figure}
         So we can formally define the set of free states as
         $\mathcal{F}:=\{\rho\in\mathbb{C}^{2}|(\rho\otimes\sigma)\in AbSEP, \forall\sigma\in\mathcal{F}\}$. Geometrically it means the free states form a sphere of radius $\simeq 0.198$ concentric with the Bloch sphere for qubit state space ({\it see Fig. \ref{fig5} }). Conversely, the set of resource states is $\mathcal{\bar{F}}:= \mathcal{S}(\mathbb{C}^{2})/\mathcal{F}$, where $\mathcal{S}(\mathbb{C}^{2})$ is the total qubit state space.  As the action of any $\mathcal{U}_{g}$ on the product of two states in $\mathcal{F}$ can not entangle it, we can characterize any entropy non-decreasing operation $\mathcal{ENDO}$ as a free operation on the set of free states. Formally $\mathcal{ENDO}(\mathcal{F})\to\mathcal{F}$. However, as the action of swap operator on the product of two free states can not make them entangled, the general class of free operations is slightly larger than $\mathcal{ENDO}$, i.e., $\mathcal{ENDO}\subset \Lambda_{free}$.\par
		Among the set of resource states $\mathcal{\bar{F}}$, the von-Neumann entropy can be characterized as a monotonic measure of the resource, i.e., $\xi$ is more resourceful than $\eta$ {\it iff} $ S(\xi)\leq S(\eta)$.\par
		Now Eq. (\ref{resource}) shows that, $\forall\rho \in \mathcal{F}, \exists~\xi \in \mathcal{\bar{F}}$, such that $\rho\otimes\xi$ can be entangled under proper choice of global unitary.
       
         \section{Conclusion}
        We have numerically characterized several classes of classically correlated and entangled bipartite state with locally thermal marginals, which can not be prepared from two same or different temperature thermal states under global unitary operations. Then the natural question arises as to what kind of correlated states can be prepared under these conditions. We have derived a criterion on the spectrum of such correlated states. We have also shown that any entangled state can be prepared from a bipartite product state if and only if it can be prepared from two thermal qubits. In continuation, a parametric class of entangled states has been proposed, with locally thermal marginals, which can always be prepared from two same or different thermal states under unitary evolution. As an application, it is possible to prepare a sustained entanglement between two spatially separated labs with different local temperatures by means of a teleportation protocol. However it imposes a bound on the difference between local temperatures. Furthermore, the upper bound on the temperature of thermal qubits to make them entangled, as proposed in \cite{huber'njp}, can be superseded by using two different thermal states, one at temperature higher than the bound and the other lower. The set of states for which the product of any two of them can not be entangled under global unitary have been classified as free states, where as the complementary qubit state space denotes the potential resources. Hence, our result can also be explained from this resource theoretic perspective, as given a free state one can always have a resource state for which the action of the global unitary on their joint product state can entangle them. More generally, we have derived a dimension dependent critical temperature below which two copies of any $d-$dimensional thermal state can be entangled in $2\times d-$dimension. However, this bound depends upon the proposed protocol. So one can investigate further to give an optimal dimension dependent temperature bound for thermal states to make them entangled. Our result indicates that the inherent structure of quantum evolution restricts the creation of correlations between thermal states, and it is interesting to establish a connection between these impossibilities to the generalized laws of quantum thermodynamics.          %In this work several classes of entangled and classically correlated states have discussed which are forbidden to prepare from two copies of thermal qubits. For more generality there is no restriction on the temperatures of initial thermal qubits. The correlated states are assumed to be locally thermal in nature, to preserve the correlation they acquire. A criterion is derived depending upon the spectrum of correlated states for which they can be prepared from two copies of identical thermal qubit. It has also been shown that the entangled states which can't be prepared from two thermal qubits, are forbidden to prepare from Separable but not absolutely separable product states. A general form of locally thermal entangled state has designed, which is allowed to prepare from two arbitrary thermal qubits and leads to share a sustain entanglement between two distant labs. However the temperature difference for the marginals of final entangled state can't be arbitrarily large, a numerical bound is estimated for this difference. But it is interesting to detect a physical principle responsible for this upper bound. Furthermore the temperature bound for thermal qubits as proposed in \cite{huber'njp} can be overcome by using two different thermal states. A dimension dependent temperature limit is obtained below which two copies of any $d-$dimensional thermal state can be entangled in $2\times d-$dimension. However, this bound is optimal depending upon the protocol. So it will be interesting to give an optimal dimension dependent temperature bound for thermal states to make them entangled.
         
         \section{Acknowledgment}
          M.A. would like to acknowledge the CSIR project 09/093(0170)/2016-EMR-I for financial support.
                       
         \begin{widetext}
         \section{Appendix}
         \subsection{Classical Correlation: CASE I}
         If $\lambda_{2}=\lambda_{3}\ge \lambda_{4}$, then,\\
                  	 $(1-p)e^{\beta E} \ge (1+ pe^{\beta E})
         	 \implies e^{\beta E}-1\ge 2pe^{\beta E}
         	 \implies p\le \frac{1-e^{-\beta E}}{2}$.\\
         	For other condition the degenerate eigenvalue will be the lowest one, hence the correlation can't be prepared from the product of two thermal states with same temperature. Then let's consider two different thermal states $(s,1-s)$ and $(t,1-t)$ with $s,t\ge\frac{1}{2}$. We also assume $s\ge t$ without loss of generality. Preservation of global spectrum under unitary demands, $st\ge s(1-t)\ge t(1-s)=(1-t)(1-s)$, where the last equality gives, {\it either} $s=1$, {\it or} $t=\frac{1}{2}$.\\
         	For $s=1$ the degenerate lowest eigenspace will be {\it null}, hence, $(1-p)e^{\beta E}=0\implies p=1$ (as $\beta E\to-\infty$ is impossible). Under this condition, $\frac{t}{1-t}=\frac{\lambda_{1}}{\lambda_{4}}|_{p=1}=e^{\beta E}$. Hence, the state (4) with $p=1$ can be prepared from two different thermal states of temperature $\frac{1}{\beta E}$ and zero, under the action of a trivial C-NOT gate by taking them {\it control} and {\it target} respectively.
         	\\For $t=\frac{1}{2}$, the first two eigenvalues of the initial product state will be identical, which gives, $(p+e^{\beta E})e^{\beta E}=(1+pe^{\beta E})\implies \beta\to0$. Hence it is possible to prepare a particular type of correlated state $\beta=0$. Therefore, $\frac{s}{1-s}=\frac{1+p}{1-p}$, which implies the temperature parameter for the initial states are, $\beta_{s}E_{s}=\ln(\frac{1+p}{1-p})$ and $\beta_{t}E_{t}=0$. As a result a single parameter $(p)$ correlated state can be prepared by fixing the other $(\beta)$ as zero, by taking one of the initial state as $\frac{\mathbb{I}}{2}$, i.e., a trivial one.
         	 \subsection{Classical Correlation: CASE II}
         	 The local marginals corresponding to the correlated state  (5) can be written as $\rho_A=\frac{1}{Z}|0\rangle\langle0|+\frac{e^{-\beta E}}{Z}|1\rangle\langle1|$ and $\rho_B=\frac{p+(1-p)e^{-\beta E}}{Z}|0\rangle\langle0|+\frac{(1-p)+pe^{-\beta E}}{Z}|1\rangle\langle1|$. Although $\rho_A$ is thermal for any arbitrary values of the parameters $\{p,\beta E\}$, $\rho_B$ will be thermal for\\
         	 $\frac{p+(1-p)e^{-\beta E}}{Z}\ge\frac{(1-p)+pe^{-\beta E}}{Z}\\
         	 \implies$ {\it either,} $p\ge \frac{1}{2}$ and $e^{-\beta E}\le 1$, {\it or,} $p\le\frac{1}{2}$ and $e^{-\beta E}\ge 1$.\\
         	 So, it is obvious that the marginals can't be locally thermal if $p\le\frac{1}{2}$.\\
         	  Now, for the region $p\ge \frac{1}{2}$, if $ \lambda_3\ge\lambda_2$ ,then
         	  \begin{eqnarray}
         	  	\begin{aligned}
         	  	p&\ge(1-p)e^{\beta E}\nonumber\\
         	  	\beta E&\le \ln(\frac{p}{1-p})
         	  	\end{aligned}
         	  \end{eqnarray}
         	   We choose product of two different thermal states $\tau_{\beta_s}=s|0\rangle\langle0|+(1-s)|1\rangle\langle1|$ and $\tau_{\beta_t}=t|0\rangle\langle0|+(1-t)|1\rangle\langle1|$ to prepare the final correlation in a restricted class. Here, $s,t\ge\frac{1}{2}$ and without loss of any generality we choose $s\ge t$. Equating the spectrum corresponding to the initial product states with the global one, we obtain, $\frac{s}{1-s}=\frac{p}{1-p}\implies\beta_sE=\ln(\frac{p}{1-p})$ and $\frac{t}{1-t}=e^{\beta E}\implies\beta_tE=\beta E$.
         	   \\For the other region, where $p\ge\frac{1}{2}$ but $\beta E\ge \ln(\frac{p}{1-p})$, the eigen values will be ordered as, $\lambda_{1}\ge\lambda_{2}\ge\lambda_{3}\ge\lambda_{4}$. Again the same analysis as above shows, $\beta_tE=\ln(\frac{p}{1-p})$ and $\beta_sE=\beta E$.
         	  \subsection{Classical Correlation: CASE III}
         	  For the first region, where $\lambda_{+}\ge\lambda_{2}=\lambda_{3}\ge\lambda_{-}$, then the final correlated state can be prepared from two same temperature thermal states. The additional condition from  Theorem 1 demands
         	  \begin{eqnarray}
         	  \begin{aligned}
         	  \lambda_{2}^{2}&=\lambda_{+}.\lambda_{-}\nonumber\\
         	  (\frac{qe^{-\beta E}}{(1+e^{-\beta E})^2})^2&=\frac{e^{2\beta E}(4p(1-q)+q(2-q)+2(q(1-q)+2p(1-q)-p^{2})-p^2)\cosh(\beta E))}{(1+e^{-\beta E})^4}\nonumber\\
         	  \frac{e^{\beta E}(p^2-(2p+q)(1-q))}{(1+e^{-\beta E})^2}&=0\\
         	  p^2&=2p-2pq+q-q^2\\
         	  (p+q)^{2}&=(2p+q)
         	  \end{aligned}
            	  \end{eqnarray}
            	  But $p+q\le1$, i.e., the last equality is impossible to satisfy except for $q=1, p=0$. As a result the state will be product in that case.\\
            	  For the second region, $\lambda_{+}\ge\lambda_{-}\ge\lambda_{2}=\lambda_{3}$. We consider two different thermal states as stated in earlier subsection. Degeneracy in the lowest eigen values implies,$(1-s)t=(1-s)(1-t)$, as a consequence, {\it either,} $s=1$, {\it or,}$t=\frac{1}{2}$.\\
            	  Now, for $s=1,~~ \lambda_{2}=\lambda_{3}=0 \implies q=0 (e^{-\beta E}\neq0).$ Therefore, in that case, initial two product states can be chosen as, $|0\rangle$ and another thermal state of temperature $\beta_{s}E=\ln(\frac{\lambda_{+}}{\lambda_{-}})$.
            	  \\Now, for $t=\frac{1}{2}, \lambda_{+}=\lambda_{-}$, as a result,\\
            	  \begin{eqnarray}
            	  \begin{aligned}
            	  	 1+4((1-p-q)^2-2)e^{-\beta E}+e^{-2\beta E}&=0\nonumber\\
            	  	 \frac{e^{\beta E}+e^{-\beta E}}{2}&=2((1-p-q)^2-2)\\
               	 \end{aligned}
               	  \end{eqnarray}
            	 The r.h.s of the above equation is negative, whereas the l.h.s is positive which leads to a contradiction.
            	 \subsection{Possibility of Entanglement}
            The action of unitary maps the probabilities $\{p^{2}\ge p(1-p)=p(1-p)\ge (1-p)^{2}\}$ to the entangled basis given by,$\{|\phi_{00}\rangle,|\phi_{01}\rangle,|\phi_{10}\rangle,|\phi_{11}\rangle\}$, as mentioned earlier, where the coefficients of these states will follow $a^{2}\ge b^{2}$. Now, we will derive two different inequalities to achieve the thermal conditions of marginals.
                     \begin{eqnarray}
            \begin{aligned}
            	p^{2}(a^{2}-b^{2})&\ge (1-p)^{2}(a^{2}-b^{2})\nonumber\\
            p^{2} a^{2}+p(1-p)+(1-p)^{2}b^{2}&\ge p^{2} b^{2}+p(1-p)+(1-p)^{2}a^{2}\\
             p^{2} a^{2}+p(1-p)(a^{2}+b^{2})+(1-p)^{2}b^{2}&\ge p^{2} b^{2}+p(1-p)(a^{2}+b^{2})+(1-p)^{2}a^{2}\\
             p^{2} a^{2}+p(1-p)(a^{2}+b^{2})+(1-p)^{2}b^{2}&\ge\frac{1}{2}.\\
            \end{aligned}
            \end{eqnarray}
            The last inequality shows the marginals of final entangled state will be thermal for the transformations, $\mathcal{U}:|ij\rangle \to |\phi_{ij}\rangle$ or,  $|ij\rangle \to |\phi_{ji}\rangle$, where $i,j\in\{0,1\}$.\\
            Again, for $p\ge \frac{1}{2}$,
            \begin{eqnarray}
            \begin{aligned}
            p^{2}+(1-p)^{2}&\ge 2p(1-p)\nonumber\\
           (p^{2}+(1-p)^{2})(a^{2}-b^{2})&\ge 2p(1-p)(a^{2}-b^{2})\\
            (p^{2}+(1-p)^{2})a^{2}+ 2p(1-p)b^{2}&\ge(p^{2}+(1-p)^{2})b^{2}+ 2p(1-p)a^{2}\\
            (p^{2}+(1-p)^{2})a^{2}+ 2p(1-p)b^{2}&\ge\frac{1}{2}\\
            \end{aligned}
            \end{eqnarray}
            This inequality with the previous one guarantee that the marginals will be thermal for the transformations $\mathcal{U}:|ij\rangle \to |\phi_{\bar{i}\bar{j}}\rangle$ or, $|ij\rangle \to |\phi_{\bar{j}\bar{i}}\rangle$, for $i,j\in\{0,1\}$. Obviously it is followed from above inequalities, that other transformations with $|00\rangle \not\to |\phi_{00}\rangle$ can't produce locally thermal marginals.
            	          	\end{widetext}

\begin{thebibliography}{99}
            \bibitem{resource} Fernando G.S.L. Brand$\tilde{a}$o {\it et. al.}, \emph{Resource Theory of Quantum States Out of Thermal Equilibrium}, \href{https://journals.aps.org/prl/abstract/10.1103/PhysRevLett.111.250404}{Phys. Rev. Lett. {\bf 111}, 250404 (2013)}
            
            \bibitem{secondlaw} Fernando Brand$\tilde{a}$o {\it et. al.}, \emph{The Second Laws of Thermodynamics}, \href{https://www.pnas.org/content/112/11/3275}{PNAS 112 (11) 3275-3279 (2015)}
            
            \bibitem{apriori} S. Popescu, A.J. Short and A. Winter, \emph{Entanglement and the foundations of statistical mechanics}, \href{https://www.nature.com/articles/nphys444}{Nature Physics {\bf 2} 754-756 (2006)}
            
            \bibitem{popescuNAT} P. Skrzypczyk, A. J. Short and S. Popescu, \emph{Work extraction and thermodynamics for individual quantum systems}, 
            \href{http://www.nature.com/ncomms/2014/140627/ncomms5185/abs/ncomms5185.html}{Nat. Comm. {\bf 5}, 4185 (2014)}
                       
         	\bibitem{allahavardyan} A.E. Allahverdyan, R.Balian and Th.M. Nieuwenhuizen,\emph{Maximal work extraction from finite quantum systems},\href{http://iopscience.iop.org/article/10.1209/epl/i2004-10101-2/meta}{EPL (Europhysics Letter) {\bf 67}, Number 4 (2004)}
         	
         	\bibitem{alicki} R. Alicki and M. Fannes, \emph{Entanglement boost for extractable work from ensembles of quantum batteries}, \href{http://dx.doi.org/10.1103/PhysRevE.87.042123}{Phys. Rev. E , {\bf 87} 042123 (2013)}
         	
         	\bibitem{franco_engine} H.T. Quan, Yu-xi Liu, C.P. Sun and Franco Nori, \emph{Quantum thermodynamic cycles and quantum heat engines
         	}, \href{https://journals.aps.org/pre/abstract/10.1103/PhysRevE.76.031105}{Phys. Rev. E , {\bf 76} 031105 (2007)}
         
         	\bibitem{refrigerator} N. Linden, S. Popescu and P. Skrzypczyk, \emph{How Small Can Thermal Machines Be? The Smallest Possible Refrigerator}, \href{https://journals.aps.org/prl/abstract/10.1103/PhysRevLett.105.130401}{Phys. Rev. Lett. , {\bf 105} 130401 (2010)}
         	
         	\bibitem{chargingbattery} N. Linden, S. Popescu and P. Skrzypczyk, \emph{How Small Can Thermal Machines Be? The Smallest Possible Refrigerator}, \href{https://journals.aps.org/prl/abstract/10.1103/PhysRevLett.105.130401}{Phys. Rev. Lett. , {\bf 105} 130401 (2010)}
         	
         	\bibitem{solarcellAlicki} R.Alicki, D. Gelbwaser-Kimovsky and A. Jenkins, \emph{A thermodynamic cycle for the solar cell}, \href{https://www.sciencedirect.com/science/article/pii/S0003491617300039?via]]]]3Dihub}{Phys. Rev. Lett. , {\bf 105} 130401 (2010)} 
         	 
        	\bibitem{bera'NAT} M.N. Bera, A. Riera, M. Lewenstein and A. Winter,\emph{Generalized laws of thermodynamics in the presence of correlations}, \href{https://www.nature.com/articles/s41467-017-02370-x}{Nature Communications {\bf 8}, 2180 (2017)}.
         	          	
        	\bibitem{bera'book} M.N. Bera, M. Lewenstein and A. Winter,\emph{Thermodynamics from Information}, \href{https://arxiv.org/abs/1805.10282}{arXiv:quant-ph 1805.1280}.
        	
        	\bibitem{horo'prl} J. Oppenheim, M. Horodecki, P. Horodecki and R. Horodecki, \emph{Thermodynamical Approach to Quantifying Quantum Correlations}, \href{https://journals.aps.org/prl/abstract/10.1103/PhysRevLett.89.180402}{Phys. rev. Lett. {\bf 89}, 180402 (2002)}
        	       	
        	\bibitem{huberPRX} M. Perarnau-Llobet {\it et. al.},\emph{Extractable Work from Correlations}, \href{https://journals.aps.org/prx/abstract/10.1103/PhysRevX.5.041011}{Phys. Rev. X{\bf 5}, 041011 (2015)}
        	
        	\bibitem{manikPRE} A. Mukherjee, A.Roy, S.S.Bhattacharya and M. Banik,\emph{Presence of quantum correlations results in a nonvanishing ergotropic gap}, \href{https://journals.aps.org/pre/abstract/10.1103/PhysRevE.93.052140}{Phys. Rev. E{\bf 93}, 052140 (2016)}
        	
        	\bibitem{our} M. Alimuddin, T. Guha and P. Parashar,\emph{Bound on Ergotropic Gap for Bipartite Separable States}, \href{https://arxiv.org/abs/1902.04869}{arXiv:quant-ph 1902.4869}.
        	
        	\bibitem{huber'njp} M. Huber {\it et. al.},\emph{Thermodynamic cost of creating correlations}, \href{https://iopscience.iop.org/article/10.1088/1367-2630/17/6/065008/meta}{ New Journal of Physics, Volume 17, June 2015}
        	
         	\bibitem{teleportation} C.H. Bennett {\it et. al.},\emph{Teleporting an unknown quantum state via dual classical and Einstein-Podolsky-Rosen channels}, \href{https://journals.aps.org/prl/pdf/10.1103/PhysRevLett.70.1895}{Phys. Rev. Lett. {\bf 70}, 1895 (1993)}
         	
         	\bibitem{densecoding} C.H. Bennett and S.J.Wiesner,\emph{Communication via one- and two-particle operators on Einstein-Podolsky-Rosen states}, \href{https://journals.aps.org/prl/pdf/10.1103/PhysRevLett.69.2881}{Phys. Rev. Lett. {\bf 69}, 2881 (1992)}.
         	
         	\bibitem{ekart} A. Ekart, \emph{Quantum cryptography based on Bell’s theorem}, \href{https://journals.aps.org/prl/pdf/10.1103/PhysRevLett.67.661}{Phys. Rev. Lett. {\bf 67}, 661-663 (1991)}
         	         	
         	\bibitem{bayesian} M. Banik {\it et. al.},\emph{Bayesian Games, Social Welfare Solutions and Quantum Entanglement},\href{https://arxiv.org/abs/1703.02773}{arXiv:quant-ph 1703.02773}
         	
         	\bibitem{ALC} S. S. Bhattacharya {\it et. al.}, \emph{Supremacy of quantum theory over supra-quantum models of communication}, \href{https://arxiv.org/abs/1806.09474}{arXiv:quant-ph 1806.09474}
         	   
         	\bibitem{ghosh16} Prathik Cherian J, S. Chakraborty and S. Ghosh, \emph{On thermalization of two-level quantum systems},\href{https://arxiv.org/abs/1604.04998}{arXiv:quant-ph 1604.04998}
         	
         	\bibitem{thermalbath} T. Guha {\it et. al.}, \emph{Thermodynamics of local baths in the context of work extraction},\href{https://arxiv.org/abs/1708.09818}{arXiv:quant-ph 1708.09818}     	
         	
         	\bibitem{res_ent} V. Vedral {\it et. al.}, \emph{Quantifying Entanglement},\href{https://journals.aps.org/prl/abstract/10.1103/PhysRevLett.78.2275}{Phys. Rev. Lett. {\bf 78}, 2275 }
         	
         	\bibitem{res_ent_1} M. A. Nielsen, \emph{Continuity bounds for entanglement},\href{https://journals.aps.org/pra/abstract/10.1103/PhysRevA.61.064301}{Phys. Rev. A {\bf 61}, 064301 }
         	
         	\bibitem{res_coh} J.\AA berg, \emph{Quantifying Superposition}, \href{https://arxiv.org/abs/quant-ph/0612146}{arXiv:quant-ph/0612146}
         	
         	\bibitem{res_pure} A. Streltsov {\it et. al.}, \emph{Maximal Coherence and the Resource Theory of Purity}, \href{https://iopscience.iop.org/article/10.1088/1367-2630/aac484/meta}{New J. Phys. {\bf 20}, 053058 (2018)}
         	
         	\bibitem{res_channel} Zi-Wen Liu and A. Winter, \emph{Resource theories of quantum channels and the universal role of resource erasure}, \href{https://arxiv.org/abs/1904.04201}{arXiv:quant-ph1904.04201}
         	
         	\bibitem{res_context} B. Amaral, \emph{Resource Theory of Contextuality}, \href{https://arxiv.org/abs/1904.04182}{arXiv:quant-ph 1904.04182}
         	
         	\bibitem{lenard} A.Lenard, \emph{Thermodynamical proof of the Gibbs formula for elementary quantum systems},\href{https://doi.org/10.1007/BF01011769}{Journal of Statistical Physics, Volume 19, Issue 6, pp 576-586 (1978)}
         	
         	\bibitem{skrzypzyckPRE} P. Skrzypczyk, R. Silva and N. Brunner, \emph{Passivity, complete passivity, and virtual temperatures},\href{https://journals.aps.org/pre/abstract/10.1103/PhysRevE.91.052133}{Phys. Rev. E {\bf 91} 052133 (2015)}
         	
         	\bibitem{ppt'PLA} M. Horodecki, P. Horodecki and R. Horodecki,\emph{Separability of mixed states: necessary and sufficient conditions},\href{https://www.sciencedirect.com/science/article/pii/S0375960196007062}{Physics Letters A, Volume 223, Issues 1–2 (1996)}
         	
         	\bibitem{ref} G. Vidal,\emph{Entanglement of Pure States for a Single Copy},\href{https://journals.aps.org/prl/abstract/10.1103/PhysRevLett.83.1046}{Phys. Rev. Lett. {\bf 83}, 1046 (1999)}
         	
         	\bibitem{wooters'PRL} W.K. Wootters,\emph{Entanglement of Formation of an Arbitrary State of Two Qubits}, \href{https://journals.aps.org/prl/abstract/10.1103/PhysRevLett.80.2245}{Phys. Rev. Lett. {\bf 80}, 2245 (1998)}
         
         \bibitem{marek} M. Kus and K. $\dot{Z}$yczkowski, \emph{Geometry of entangled states}, \href{https://journals.aps.org/pra/abstract/10.1103/PhysRevA.63.032307}{Phys. Rev. A {\bf 63}, 032307 (2001)}
         
         \bibitem{knill} E. Knill,\emph{Separability from Spectrum}, \href{https://qig.itp.uni-hannover.de/qiproblems/Separability_from_spectrum}{ http://qig.itp.uni-hannover.de/qiproblems/15 (2003)}
         
         \bibitem{Johnston'13} N. Johnston,\emph{Separability from spectrum for qubit-qudit states}, \href{https://journals.aps.org/pra/abstract/10.1103/PhysRevA.88.062330}{Phys. Rev. A {\bf 88}, 062330 (2013)}
         
         \bibitem{comment}{For our discussion we will treat $|0\rangle$ and $\frac{\mathbb{I}}{2}$ as trivial thermal states which represents $T=0K$ and $T\to\infty$ respectively. By mentioning allowed correlation we mean those which can be prepared from other quantum states except these two.}
         
         \bibitem{HuberPRE'15} D. E. Bruschi {\it et. al.}, \emph{Thermodynamics of creating correlations: Limitations and optimal protocols}, \href{https://journals.aps.org/pre/abstract/10.1103/PhysRevE.91.032118}{Phys. Rev. E {\bf 91}, 032118 (2015)}
         \end{thebibliography}
         	      \end{document}